# Magnonic crystals — prospective structures for shaping spin waves in nanoscale

J. Rychly, P. Gruszecki, M. Mruczkiewicz, J.W. Kłos, S. Mamica, and M. Krawczyk

*Faculty of Physics, Adam Mickiewicz University in Poznan, 85 Umultowska, Poznań 61-614, Poland*
E-mail: krawczyk@amu.edu.pl

We have investigated theoretically band structure of spin waves in magnonic crystals with periodicity in one- (1D), two- (2D) and three-dimensions (3D). We have solved Landau–Lifshitz equation with the use of plane wave method, finite element method in frequency domain and micromagnetic simulations in time domain to find the dynamics of spin waves and spectrum of their eigenmodes. The spin wave spectra were calculated in linear approximation. In this paper we show usefulness of these methods in calculations of various types of spin waves. We demonstrate the surface character of the Damon–Eshbach spin wave in 1D magnonic crystals and change of its surface localization with the band number and wavenumber in the first Brillouin zone. The surface property of the spin wave excitation is further exploited by covering plate of the magnonic crystal with conductor. The band structure in 2D magnonic crystals is complex due to additional spatial inhomogeneity introduced by the demagnetizing field. This modifies spin wave dispersion, makes the band structure of magnonic crystals strongly dependent on shape of the inclusions and type of the lattice. The inhomogeneity of the internal magnetic field becomes unimportant for magnonic crystals with small lattice constant, where exchange interactions dominate. For 3D magnonic crystals, characterized by small lattice constant, wide magnonic band gap is found. We show that the spatial distribution of different materials in magnonic crystals can be explored for tailored effective damping of spin waves.

## Introduction

Material patterning is one of the major factors used for controlling propagation of waves with short wavelengths in solids. The use of periodic patterning for controlling the wave dynamics has increased significantly since the discovery of photonic crystals in 1987 [1,2]. Since that time periodicity has been extensively used for molding the flow of electromagnetic waves in the range from microwaves to optical wavelengths [3], which led to the discovery of new materials with properties unheard of in nature [4]. Ideas developed in photonics were transferred to other types of excitations, such as phonons [5], plasmons [6] and also spin waves (SWs) [7].

Periodicity was used for controlling SWs excitations in ferromagnetic materials already in 1970s [8,9]. The transmission of magnetostatic waves was tailored by periodic distribution of the saturation magnetization, realized by ion implantation [10], a regular lattice of etched grooves in a magnetic dielectric, periodic modulation by metallic stripes or dots on top of a ferromagnetic film [11], or periodic perturbation of the magnetic field [12]. Due to limitations in fabrication and technology the investigations were limited to large structures, in which the dipolar interaction prevailed over the exchange interaction. The discovery of photonic crystals renewed the interest in magnetic periodic structures, inspiring many new ideas, providing abundant new physics, and pushing magnetization dynamics studies in unexplored directions. The concept of *magnonic crystals* (MCs) was proposed as a SW counterpart of photonic crystals [13–16]. MCs are magnetic structures with periodic distribution of the constituent materials or periodic modulation of some magnetic parameters (e.g. saturation magnetization, exchange interactions or magnetocrystalline anisotropy), or other parameters relevant to the propagation of SWs, such as external magnetic field, film thickness, stress or a surrounding of the ferromagnetic film.

In any periodic medium the eigensolutions of the wave equation in the linear regime fulfill Bloch's theorem and, regardless of the type of excitation, form a band structure in the frequency–wave vector space. The same applies to the magnonic (i.e., spin-wave) band structure. Nevertheless, the details of the band structure can be derived mostly from numerical calculations. Usually the magnonic band structure is studied in terms of its sensitivity to: structural changes of the MC, modifications of material parameters or application of external fields.

SWs have a complex dispersion relation even in homogeneous thin film [16] which is substantially different from the dispersion of SWs in a bulk material. The dispersion depends on the direction and magnitude of the wave vector, the relative strength of short-range exchange interaction with respect to the strength of long-range dipolar interaction, the external shape of the sample, the magnitude of the applied static magnetic field and its orientation in relation to the direction of SWs propagation. Also the magnetocrystalline anisotropy and magnetoelastic effects contribute to SWs dynamics. This means that the SW band structure of MCs will be influenced by many additional factors, both intrinsic and external ones, apart from those which are typical for the other types of artificial crystals. This makes MCs an intriguing system for scientific studies and the main subject of research in the field of *magnonics* [7].

This paper is dedicated to present various spin wave band structure which can be realized in MCs to demonstrate interesting properties of SWs resulting from the periodicity. We present the results of calculations for one-dimensional (1D), two-dimensional (2D) and tree-dimensional (3D) MCs obtained with different numerical methods. All spectra are obtained by solving Landau–Lifshitz (LL) equation. The plane wave method, finite element method in the frequency domain and micromagnetic simulations based on finite difference method in the time domain are demonstrated to be complementary for calculations of the SW spectra in MCs.

## Model and wave equation

The dynamics of SWs can be analyzed using two main theoretical approaches [18]. One uses the discrete lattice model based on Heisenberg Hamiltonian with atomic structure of the ferromagnetic material directly taken into account. The other is based on solutions of the LL equation defined in continuous medium which describes precession of classical magnetic moments in effective magnetic field. The later approach is more suitable for systems with complex geometry in nano- and larger scales, where local parameters (exchange integral, spin) can be expressed by macroscopic parameters (exchange length and magnetization saturation) [19]. The typical spatial resolution of nanolitography techniques is order of magnitude larger than interatomic distances. This justify use of an approach based on LL equation to these structures, which are widely fabricated nowadays [20].

We solve the LL equation, i.e., the equation of motion for the magnetization vector $\mathbf{M}(\mathbf{r},t)$:

$$\frac{\partial \mathbf{M}(\mathbf{r},t)}{\partial t} = -\gamma\mu_0 \mathbf{M}(\mathbf{r},t) \times \mathbf{H}_{\text{eff}}(\mathbf{r},t) + \frac{\alpha\gamma\mu_0}{M_S} \mathbf{M}(\mathbf{r},t) \times [\mathbf{M}(\mathbf{r},t) \times \mathbf{H}_{\text{eff}}(\mathbf{r},t)], \quad (1)$$

where $\gamma$ is gyromagnetic ratio, $\mathbf{H}_{\text{eff}}(\mathbf{r},t)$ denotes effective magnetic field, $\mu_0$ is permeability of vacuum, $\mathbf{r}$ is position vector, $t$ is time and $M_S$ is saturation magnetization. Relaxation processes are described by the last term on the right-hand side of the Eq. (1). We assume that MC is saturated, i.e., a collinear static magnetization in all the investigated structures is assumed. In this paper we use a coordinate system with the $z$ axis being the direction of the external magnetic field and also the direction of the static magnetization in the saturated state.

In the case of small disturbance of the magnetization from its equilibrium orientation, linear SWs are generated and the calculations can be conducted in linear approximation. In this case the component of the magnetization vector along equilibrium direction ($z$ axis) is constant in time [$\mathbf{M}(\mathbf{r},t) = M_z(\mathbf{r})\hat{z} + \mathbf{m}(\mathbf{r},t)$] and can be approximated by spontaneous magnetization $M_z(\mathbf{r}) \approx M_S(\mathbf{r})$. This assumption requires much larger magnitude of $M_z$ than those of the perpendicular components: $|\mathbf{m}(\mathbf{r},t)| \ll M_S(\mathbf{r})$, where $\mathbf{m}(\mathbf{r},t)$ is a two-dimensional dynamic vector lying in the $(x, y)$ plane: [$\mathbf{m}(\mathbf{r}, t) = m_x(\mathbf{r}, t)\hat{x} + m_y(\mathbf{r}, t)\hat{y}$].

The effective magnetic field is assumed to be the sum of three terms: $\mathbf{H}_{\text{eff}} = \mathbf{H}_0 + \mathbf{H}_{\text{ms}} + \mathbf{H}_{\text{ex}}$, where $\mathbf{H}_0$ is the external static magnetic field; $\mathbf{H}_{\text{ms}}$ is the magnetostatic field with two components: static demagnetizing field $\mathbf{H}_{\text{dem}}(\mathbf{r})$ and dynamic components $\mathbf{h}(\mathbf{r}, t)$ that are perpendicular to $\mathbf{H}_0$: $\mathbf{H}_{\text{ms}} = [h_x, h_y, H_{\text{dem}}]$; $\mathbf{H}_{\text{ex}}$ is the exchange field which can be formally defined in saturation using linear space dependent exchange operator $\hat{\mathbf{H}}_{\text{ex}}$: $\mathbf{H}_{\text{ex}}(\mathbf{r}) = \hat{\mathbf{H}}_{\text{ex}}(\mathbf{r})\mathbf{m}(\mathbf{r})$. We neglect the contribution of the magnetic anisotropy.

In magnetostatic approximation the magnetostatic field can be expressed as a gradient of the scalar magnetostatic potential: $\mathbf{H}_{\text{ms}} = -\nabla\varphi$, thus the curl of magnetostatic field (the one of the Maxwell equations) is always equal to zero: $\nabla \times \mathbf{H}_{\text{ms}} = 0$. In order to find the dynamic components of the demagnetizing field $(h_x, h_y)$, we need to solve the Gauss equation: $\nabla \cdot [\mathbf{H}_{\text{ms}}(\mathbf{r}) + \mathbf{M}(\mathbf{r})] = 0$. Putting to the Gauss equation magnetostatic field written as the gradient of $\varphi$ we receive equation:

$$\nabla^2 \varphi(\mathbf{r}) = \frac{\partial m_x(\mathbf{r})}{\partial x} + \frac{\partial m_y(\mathbf{r})}{\partial y} + \frac{\partial M_S(\mathbf{r})}{\partial z}. \quad (2)$$

This equation defines the relation between magnetostatic potential (and magnetostatic field) and the magnetization. This shall be solved together with Eq. (1).

In Eq. (1) with linear approximation we can separate space and time variables. Then, the time dependent equation has solutions in the form of monochromatic spin waves: $\mathbf{m}(\mathbf{r},t) = (m_x, m_y) \propto e^{-i\omega t}$ (similar for dynamic components of the magnetostatic field $\mathbf{h}(\mathbf{r},t) = (h_x, h_y) \propto e^{-i\omega t}$), where ω is the angular frequency of SW. Thus, the monochromatic SW is a solution of the following equations:

$$i\Omega m_x(\mathbf{r}) = H_0 m_y(\mathbf{r}) + H_{\text{dem}} m_y(\mathbf{r}) -$$
$$- M_S(\mathbf{r}) h_y(\mathbf{r}) - M_S(\mathbf{r}) \mathbf{H}_{\text{ex}}(\mathbf{r}) m_y(\mathbf{r}),$$
$$-i\Omega m_y(\mathbf{r}) = H_0 m_x(\mathbf{r}) + H_{\text{dem}} m_x(\mathbf{r}) -$$
$$- M_S(\mathbf{r}) h_x(\mathbf{r}) - M_S(\mathbf{r}) \mathbf{H}_{\text{ex}}(\mathbf{r}) m_x(\mathbf{r}), \quad (3)$$

where $\Omega = \dfrac{\omega}{\gamma \mu_0}$ is reduced angular frequency of the SW.

### Planar magnonic crystals

The simplest MC is a 1D periodic structure which has a form of magnetic multilayers [21]. These systems were intensively studied in the past and their fundamental features are well-understood using both discrete and continuous models [22]. More complex spin wave dynamics can be observed in planar structures with in-plane periodicity [23,24]. The planar geometry is the most common for the nano- and microstructures fabricated using top-down techniques (e.g. nanolithography) [20]. The obtained structures can possess long-range order of high precision, which is crucial to explore the effects of the periodicity in SW dynamics. On the other hand, the bottom-up methods [25], are effective by utilizing processes of self-organization, however they often fail in fabrication of ideal 2D and 3D periodic systems because of incontrollable defects and dislocation. Because of the mentioned above reasons, both high quality 1D and 2D MCs are fabricated mainly in the form of planar systems with in-plane dimensions much larger than thickness of the structure [26].

MCs, as any other kind of periodic medium, has anisotropic dynamics (according with symmetry of the structure) but the geometry is not the only source of the anisotropy in SW dynamics. An inhomogeneity of the static magnetization distribution induces demagnetization and exchange fields, the additional two factors which influence the SW dynamics. In MCs these magnetic fields have also periodic distribution, however they introduce additional inhomogeneity and anisotropy which can reduce symmetry of the lattice. The other factor which is inevitable for magnonic systems is an external magnetic field which can be arbitrarily oriented with respect to the crystallographic lattice axes or MC plane. The magnetic ground state (spatial distribution of the static magnetization) can be very complex in MCs at low external magnetic field, below the saturation field. The investigation of SW dynamics in such system is complicated, because of complexity of the ground state and presence of remagnetization processes. To design MCs with desired SW dispersion, the stable magnetic configuration is usually required. Strong enough magnetic field can be always used to enter system in saturation state. Under this condition, the external magnetic field determines the direction of the static magnetization, which is almost homogeneous and constant in different pieces of the same material. The magnetization dynamics can be then mostly attributed to SWs.

### Plane wave method

In this section we present an adaptation of the plane wave method (PWM) to the calculations of the SW dispersion relation in planar MCs with 2D periodicity. The general algorithm of this method is presented schematically in Fig. 1. If surface magnetic anisotropy on the surfaces of the MC is neglected and the thickness of the planar MC is small then the magnetocrystalline and dipolar pinning on its top and bottom surfaces is week. Therefore, we can assume that the magnetization is free on film's surfaces. Moreover, because the ratio of the period to thickness for the MC is small, the lowest magnonic bands will represent the modes which are not quantized across thickness of the MC. These two assumptions allow to consider the SW amplitude as constant across the slab thickness in thin planar MCs. Using this approach we will proceed with PWM, which applies 2D Fourier transform to linearized LL Eq. (3) supplemented with Gauss Eq. (2) and converts this equations into the algebraic eigenproblem. This consist first part of the algorithm presented in Fig. 1.

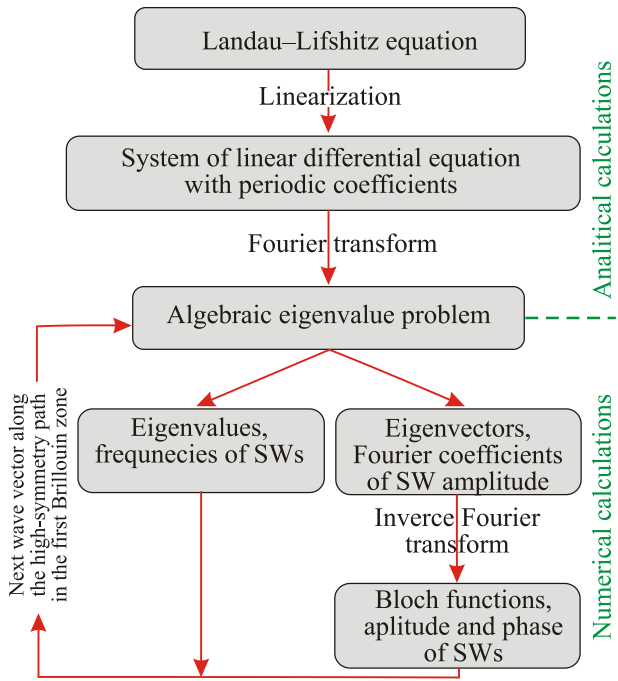

*Fig. 1.* (Color online) Algorithm of the PWM to calculate the dispersion relation of SWs and their amplitude distribution in magnonic crystal. It is split into two parts, analytical derivation of the algebraic eigenproblem and numerical solutions.

To illustrate PWM, we will discuss the solution for the system presented in Fig. 2. This is a square lattice (with lattice constant $a$) of the square dots (of side size $s$). Dots (ferromagnetic material A) and matrix (ferromagnetic material B) are saturated by the in-plane external magnetic field. The solutions of Eq. (3), which is differential equation with coefficients being periodic function of the position vector, can be written in the form of the Bloch waves:

$$\mathbf{m}(\mathbf{r}) = \tilde{\mathbf{m}}_\mathbf{k}(\mathbf{r})e^{i\mathbf{k}\cdot\mathbf{r}}, \qquad (4)$$

where $\mathbf{k} = (0, k_x, k_y)$ is the wavevector and $\tilde{\mathbf{m}}_\mathbf{k}(\mathbf{r})$ is an SWs complex amplitude with the same periodicity as the lattice of the MC, $\tilde{\mathbf{m}}_\mathbf{k}(\mathbf{r}+\mathbf{a}) = \tilde{\mathbf{m}}_\mathbf{k}(\mathbf{r})$, $\mathbf{a} = (0, a_y, a_z)$ and $a_y$ and $a_z$ are components of a two dimesional lattice vector. For a given wave vector $\mathbf{k}$, we expand the periodic factor of the Bloch function $\tilde{\mathbf{m}}_\mathbf{k}(\mathbf{r})$ and material parameters (i.e., saturation magnetization and exchange constant) in the basis of the plane waves: $e^{i\mathbf{G}\cdot\mathbf{r}}$

$$\tilde{\mathbf{m}}_\mathbf{k}(\mathbf{r}) = \sum_\mathbf{G} \mathbf{m}_\mathbf{k}(\mathbf{G})e^{i\mathbf{G}\cdot\mathbf{r}} \text{ and } M_S(\mathbf{r}) = \sum_\mathbf{G} M_S(\mathbf{G})e^{i\mathbf{G}\cdot\mathbf{r}} \quad (5)$$

where $\mathbf{G}$ is a reciprocal lattice vector, which for the square lattice takes the values: $\mathbf{G} = (0, G_y, G_z) = \frac{2\pi}{a}(0, n_y, n_z)$, with $n_y, n_z$ being integers.

Also, in Eq. (2) the magnetostatic potential can be represented as a Bloch function, and its periodic part can be expanded in Fourier series, together with $M_S$ or Bloch functions $[m_y(\mathbf{r}), m_x(\mathbf{r})]$ included in this equation. The solution of Eq. (2) for static demagnetizing field in the planar structure with piecewise constant magnetization of arbitrary orientation was derived by Kaczer *et al.* in Ref. 27.

These formulas were extended to dynamical components of the magnetization in the form of the Bloch function (4) in Ref. 28. Finally, $H_{\mathrm{dm},z}(\mathbf{r})$ and $h_x(\mathbf{r}), h_y(\mathbf{r})$ can be expressed as a function of the Fourier coefficients $M_S(\mathbf{G})$, $m_x(\mathbf{G})$ and $m_y(\mathbf{G})$ in the following form:

$$H_{\mathrm{dm}}(\mathbf{r}) = -\sum_\mathbf{G} \frac{M_S(\mathbf{G})G_z^2}{\mathbf{G}^2}\left[1 - S(\mathbf{G}, x)\right]e^{i\mathbf{G}\cdot\mathbf{r}}, \qquad (6)$$

$$h_x(\mathbf{r}) = \sum_\mathbf{G}\left[ im_y(\mathbf{G})\frac{k_y + G_y}{|\mathbf{k}+\mathbf{G}|}S(\mathbf{k}+\mathbf{G}, x) - m_x(\mathbf{G})C(\mathbf{k}+\mathbf{G}, x)e^{i(\mathbf{k}+\mathbf{G})\cdot\mathbf{r}} \right], \qquad (7)$$

$$h_y(\mathbf{r}) = \sum_\mathbf{G}\left[ im_x(\mathbf{G})\frac{k_y + G_y}{|\mathbf{k}+\mathbf{G}|}S(\mathbf{k}+\mathbf{G}, x) - \frac{m_y(\mathbf{G})(k_y+G_y)^2}{|\mathbf{k}+\mathbf{G}|^2}[1 - S(\mathbf{k}+\mathbf{G}, x)]e^{i(\mathbf{k}+\mathbf{G})\cdot\mathbf{r}} \right], \qquad (8)$$

where

$$S(\boldsymbol{\kappa}, x) = \sinh(|\boldsymbol{\kappa}|x)e^{-\frac{|\boldsymbol{\kappa}|d}{2}}, \quad C(\boldsymbol{\kappa}, x) = \cosh(|\boldsymbol{\kappa}|x)e^{-\frac{|\boldsymbol{\kappa}|d}{2}}$$

and $d$ is the MC's thickness.

In PWM we use exchange field in the form

$$\mathbf{H}_{\mathrm{ex}} = \left[\nabla\cdot\left(\frac{2A}{\mu_0 M_S^2}\right)\nabla\right]\mathbf{m}(\mathbf{r}) = \nabla\cdot l_{\mathrm{ex}}^2(\mathbf{r})\nabla\mathbf{m}(\mathbf{r}),$$

where $l_{\mathrm{ex}}(\mathbf{r}) = \sqrt{\frac{2A}{\mu_0 M_S^2}}$ is the exchange length. The expression in the square brackets is the exchange operator.

According to Bloch theorem (4) and using the expansions (5)–(8), the Eq. (3) can be transformed into the algebraic eigenproblem, which finalize the first part of the algorithm shown in Fig. 1:

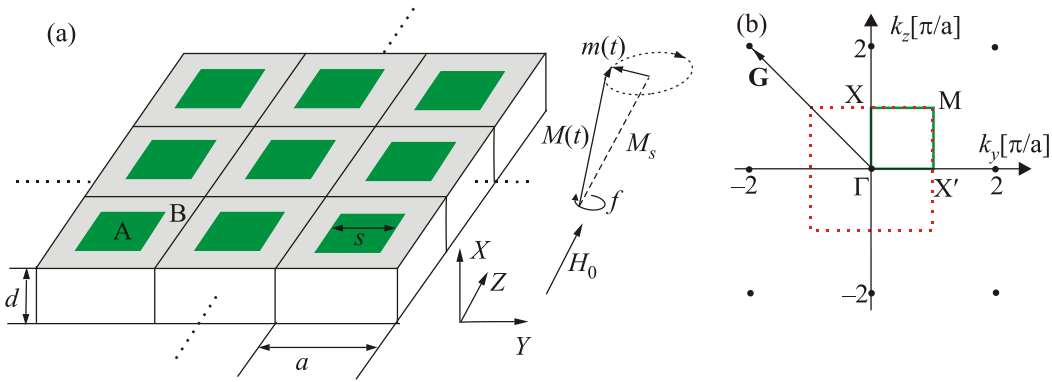

*Fig. 2.* (Color online) (a) The schema of the structure of bi-component planar magnonic crystal. Square lattice (lattice constant, $a$) of square shaped dots (with size $s$) immersed in ferromagnetic material B. The MC is saturated by external magnetic field along the z-axis. (b) Reciprocal lattice of the square lattice with the first Brillouin zone marked with red dotted line. The irreducible part of the Brillouin zone is marked with bold solid line.

$$\hat{\mathbf{M}}\mathbf{m_k} = i\Omega \mathbf{m_k}. \qquad (9)$$

The eigenvector $\mathbf{m_k} = [\mathbf{m_{k,x}}, \mathbf{m_{k,y}}]$ consists of the two sub-vectors being the finite sets (of $N$ elements) of Fourier coefficients for periodic factor of the Bloch functions:

$$\mathbf{m_{k,x}} = [m_{\mathbf{k},x}(\mathbf{G}_1), m_{\mathbf{k},x}(\mathbf{G}_2), \cdots, m_{\mathbf{k},x}(\mathbf{G}_N)] \text{ and}$$

$$\mathbf{m_{k,y}} = [m_{\mathbf{k},y}(\mathbf{G}_1), m_{\mathbf{k},y}(\mathbf{G}_2), \cdots, m_{\mathbf{k},y}(\mathbf{G}_N)].$$

The matrix $\mathbf{M}$ is a block matrix

$$\hat{\mathbf{M}} = \begin{pmatrix} M^{xx} & M^{xy} \\ M^{yx} & M^{yy} \end{pmatrix}. \qquad (10)$$

The blocks of the matrix (10) are defined as:

$$M^{xx}_{ij} = -i \frac{k_y + G_{y,j}}{|\mathbf{k}+\mathbf{G}_j|} M_S(\mathbf{G}_i - \mathbf{G}_j) S(\mathbf{k}+\mathbf{G}_j, x) = -M^{yy}_{ij},$$

$$M^{xy}_{ij} = H_0 \delta_{ij} + \sum_l (\mathbf{k}+\mathbf{G}_j)\cdot(\mathbf{k}+\mathbf{G}_l) l^2_{\text{ex}}(\mathbf{G}_l - \mathbf{G}_j) M_S(\mathbf{G}_i - \mathbf{G}_l) + \frac{(k_y + G_{y,j})^2}{|\mathbf{k}+\mathbf{G}_j|^2} M_S(\mathbf{G}_i - \mathbf{G}_j)\bigl[1 - S(\mathbf{k}+\mathbf{G}_j, x)\bigr] -$$

$$- \frac{(G_{z,i} + G_{z,j})^2}{|\mathbf{G}_i - \mathbf{G}_j|^2} M_S(\mathbf{G}_i - \mathbf{G}_j)\bigl[1 - S(\mathbf{G}_i + \mathbf{G}_j, x)\bigr],$$

$$M^{yx}_{ij} = -H_0 \delta_{ij} - \sum_l (\mathbf{k}+\mathbf{G}_j)\cdot(\mathbf{k}+\mathbf{G}_l) l^2_{\text{ex}}(\mathbf{G}_l - \mathbf{G}_j) M_S(\mathbf{G}_i - \mathbf{G}_l) - M_S(\mathbf{G}_i - \mathbf{G}_j) C(\mathbf{k}+\mathbf{G}_j, x) +$$

$$+ \frac{(G_{z,i} + G_{z,j})^2}{|\mathbf{G}_i - \mathbf{G}_j|^2} M_S(\mathbf{G}_i - \mathbf{G}_j)\bigl[1 - S(\mathbf{G}_i + \mathbf{G}_j, x)\bigr].$$

$i$ and $j$ index reciprocal lattice vectors used in the Fourier expansions. For the numerical solution of the Eq. (9), we need to limit a number of the reciprocal lattice vectors in all expansions and also calculate the coefficients of the Fourier expansion (5) of the magnetization saturation and exchange length, $M_S(\mathbf{G})$ and $l^2_{\text{ex}}(\mathbf{G})$, respectively. The general formula for the coefficients of Fourier expansion reads:

$$F(\mathbf{G}) = \frac{1}{S}\int_S F(\mathbf{r}) e^{-i\mathbf{G}\cdot\mathbf{r}} dS,$$

where $F(\mathbf{r})$ and $F(\mathbf{G})$ are periodic functions in real space describing the spatial distribution of material parameter (this is $M_S(\mathbf{r})$ or $l^2_{\text{ex}}(\mathbf{r})$ in Eq. (3)) and its Fourier expansion coefficient for reciprocal lattice vector $\mathbf{G}$ (this is $M_S(\mathbf{G})$ or $l^2_{\text{ex}}(\mathbf{G})$ in eigenvalue problem Eq. (9)), respectively. The $S$ denotes the area of the unit cell. The Fourier coefficients for MC with inclusions of the regular shapes can be analytically calculated. For instance for 2D MC with inclusions of circular and square shapes, the coefficients $F_{\text{cir}}(\mathbf{G})$ and $F_{\text{sq}}(\mathbf{G})$ take the following form:

$$F_{\text{cir}}(\mathbf{G}) = \frac{1}{S}\begin{cases} F_B + (F_A - F_B)\pi R^2 & \text{for } \mathbf{G}=0, \\ (F_A - F_B)\dfrac{2\pi R^2}{G} J_1(RG) & \text{for } \mathbf{G}\neq 0, \end{cases}$$

$$F_{\text{sq}}(\mathbf{G}) = \frac{1}{S}\begin{cases} F_B + (F_A - F_B)s^2 & \text{for } \mathbf{G}=0, \\ (F_A - F_B)\,\text{sinc}\!\left(\dfrac{G_x s}{2}\right) s^2 & \text{for } G_x \neq 0, G_y = 0, \\ (F_A - F_B)\,\text{sinc}\!\left(\dfrac{G_y s}{2}\right) s^2 & \text{for } G_x = 0, G_y = 0, \\ (F_A - F_B)\,\text{sinc}\!\left(\dfrac{G_x s}{2}\right) \text{sinc}\!\left(\dfrac{G_y s}{2}\right) s^2 & \text{for } \mathbf{G}\neq 0, \end{cases}$$

where $R$ is the radius of circular inclusions, $G$ stands for the length of the reciprocal lattice vector $\mathbf{G}$ and $J_1$ denotes the Bessel function of the first kind. $F_A$ and $F_B$ stands for the respective values of saturation magnetization in the inclusion material and host material, respectively. For the square lattice $S = a^2$ (Fig. 2(a)).

The algebraic eigenproblem (9) is solved numerically using standard routines to find eigenvalues and eigenvectors, and this constitute the second part of the algorithm. The solutions are eigenvalues $\Omega$, from which frequency $f$ of the SW modes for given $\mathbf{k}$-vector are obtained, and eigenvectors $\mathbf{m_k}$, which are used to obtain distribution of the SW amplitude in real space, $\mathbf{m_k}(\mathbf{r})$. The dispersion relation is found by solving eigenproblem (9) repetitively for successive values of the $\mathbf{k}$ along the high symmetry path in the first Brillouin zone (BZ). For square lattice this path is marked with red dotted line in Fig. 2(b).

**Magnonic band structure in thin 2D MCs**

The spectrum of MCs can be tailored by adjustment of structural and material parameters. The following parameters can be considered for system presented in Fig. 2: material composition of the matrix and inclusions, lattice constant $a$, size $s$ and shape of the inclusions, and also the thickness of the slab $d$. It is worth to notice that due to interplay between dipolar and exchange interactions, the spectrum of magnonic crystals do not scale with the size of the system, as it is in photonic systems [3]. To demonstrate basic changes in the magnonic band structure due to structural parameters, we investigate the variation of the absolute values of structural parameters but only their relative changes, even if we are interested in exploring qualitative features of the magnonic spectrum. Let's consider MCs in two different limits of sizes [29]: exchange dominated regime — for small lattice constant and dipole dominated regime — for larger patterns. In Fig. 3 we present results for of PWM calculations MC with small lattice constant ($a$ = 50 nm) for various shapes of inclusions: square, hexagonal and circular, with fixed filling fraction ($ff$ = 0.55, the filling fraction is defined as a ratio of the area occupied by inclusion to the area of the unit cell, for structure from Fig. 2 this is $ff = s^2/a^2$). The first magnonic band does not change significantly with the changes of the shape of inclusions. Also the corresponding profiles of dynamical magnetizations, which don't reproduce all details of the inclusions, are similar. In this range of sizes, the exchange interactions, which are isotropic in terms of external field direction, dominate. This property makes the MC a close counterpart of the photonic crystal and allows to deduce a lot of its features in analogy to photonics. However, the impact of dipole interactions is still noticeable in these MCs. We can observe differences in the dispersions for the equivalent crystallographic directions — $cf$, the dispersion of the first band along Γ–X and Γ–X′, parallel and perpendicular to the external magnetic field.

Figure 4 presents the dispersions for planar MCs of larger sizes (lattice constant $a$ = 400 nm), with important influence of the dipole interactions. We observe the dramatic change of the magnonic spectra for systems differing in the shape of inclusions (note that we kept the filling fraction constant for the three considered shapes). The one of the most important factors responsible for such differences in the SW spectra is the static demagnetizing filed (see the right insets in Fig. 4). This field has a form of sharp wells/peeks located at the interfaces of materials differing in magnetization saturation. The amplitude of demagnetizing field depends both on the magnetization contrast (on both sides of interface) and the orientation of the interface with respect to the direction of the external magnetic field. The later feature is shape-dependent. The strongest impact of the demagnetizing field is noticed for the system with square inclusions where long sides of squares are oriented perpendicularly to the direction of $H_0$. This induces deep and long wells of demagnetizing fields which can capture the modes and localize them. These modes (in Fig. 4(a) modes 1 and 2) are called edge modes and have frequency below the frequency of the fundamental mode.

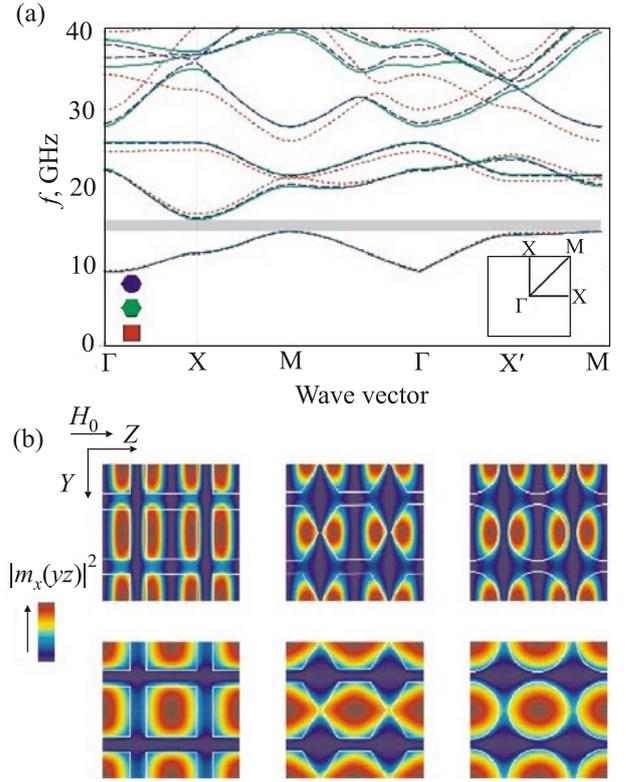

*Fig. 3.* (Color online) (a) Dispersion relation for planar MC formed by Ni inclusions embedded in Fe matrix in the small lattice constant regime ($a$ = 50 nm). The filling fraction was fixed to the value 0.55 for different shapes of inclusions. Both, the dispersions (a) and (b) the profiles of the out-of-plane component of dynamical magnetization do not show significant difference for the systems with different shapes of inclusions. We presented only the profiles for the first (bottom row) and the second (top row) band in the center of the Brillouin zone (Γ point). The thickness of the slab is 5 nm. The external field magnitude is 50 mT. The following values of the material parameters were assumed: $M_{S,\text{Fe}}$ = = 1.752·10$^{-6}$ A/m, $l_{\text{ex,Fe}}$ = 3.30 nm, and $M_{S,\text{Ni}}$ = 0.484·10$^{-6}$ A/m, $l_{\text{ex,Ni}}$ = 7.64 nm. The gyromagnetic ratio is assumed to have the same value: γ = 176 GHz/T for both materials [30].

We discussed the PWM and its adaptation to calculation of the SW dispersion relation in the planar MC. One of the assumption we have made was about homogeneity of the sample across its thickness. However, there are planar MCs which are nonuniform in out-of-plane direction, e.g. magnetic slab with an array of grooves or with magnetic dots on its surface. In these cases, this assumption can be avoided, if the slab is with tiny periodic modulation on its top surface. Then we can apply the method presented in Ref. 31 where the Walker equation [18] with modulated boundary conditions was solved in the plane wave basis to find magnonic spectrum in the magnetostatic approximation. The planar MCs of more sophisticated geometry, with competing magnetostatic and exchange interactions can be investigated using other methods. In the next two sections we demonstrate finite element method in frequency domain and micromagnetic simulations with time domain finite difference used to calculate magnonic band structure in planar 1D MC with inhomogeneity across the thickness taken into account.

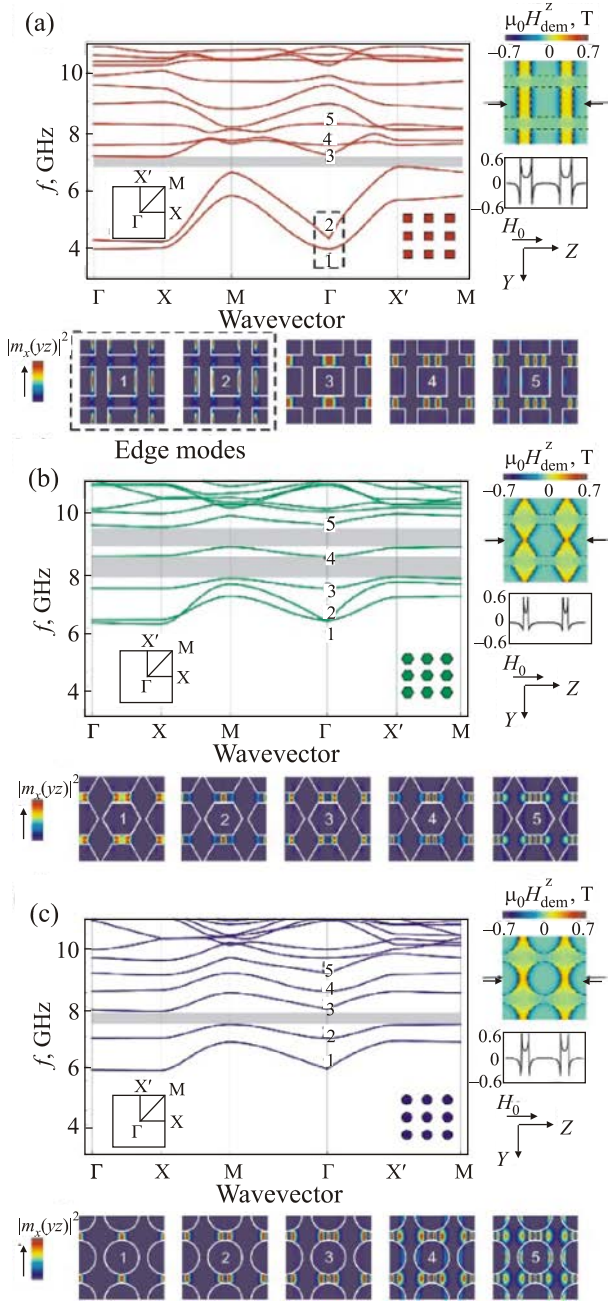

*Fig. 4.* (Color online) Dispersion relation for planar 2D MC formed by Fe inclusions of (a) square, (b) hexagonal, (c) circular shape embedded in Ni matrix in square lattice with lattice constant $a = 400$ nm. The filling fraction was fixed to the 0.55 for different shapes of inclusions. The profiles of the out-of-plane component of dynamical magnetization for five lowest modes in the center of the Brillouin zone are presented in the bottom rows in each figure. The spectra and profiles depends significantly on the shape of inclusion due to the impact of the demagnetizing field $H^z_{\text{dem}}$ (the insets on the left of each figure). The thickness of the MC is 20 nm. The external field $H_0$ along the z-axis has magnitude 100 mT. The material parameters were assumed the same as in Fig. 3 [30].

### Finite element method in frequency domain

In calculations which take into account the inhomogeneity of the magnetization across the MC thickness we considered planar 1D MCs. The schematic structure considered here is shown in Fig. 5(a). It is composed of infinitely long cobalt (Co) and permalloy (Py) stripes. Co and Py stripes are placed side by side. Their dimensions are the same: thickness $d = 30$ nm and width $a_{\text{Co}} = a_{\text{Py}} = 100$ nm, the lattice constant is $a = 200$ nm. The external magnetic field is directed along the stripes and has the magnitude $\mu_0 H_0 = 0.1$ T. The material parameters of Co and Py are as follows: Co saturation magnetization $M_{S,\text{Co}} = 1.45 \cdot 10^6$ A/m, Py saturation magnetization: $M_{S,\text{Py}} = 0.7 \cdot 10^6$ A/m; Co exchange constant: $A_{\text{Co}} = 3 \cdot 10^{-11}$ J/m and Py exchange constant: $A_{\text{Co}} = 1.1 \cdot 10^{-11}$ J/m. Studied configuration, in which external magnetic field is directed along the stripes (the z-axis), and SW propagation is perpendicular to the stripes (along the y-axis) in the film plane, is called Damon–Eshbach (DE) geometry [18]. For this geometry the static demagnetizing field equals to zero: $H_{\text{dem}} = 0$. Structure is finite in the x direction, which is a thickness of the structure.

The finite element method (FEM) is a numerical technique for finding approximate solutions to boundary value problem for partial differential equations. It uses subdivi-sion of a computational domain into smaller parts, called finite elements (the computational domain used in our computations is shown in (Fig. 5(c)). FEM encompasses simpler methods for connecting many elementary equations over small subdomains (finite elements), to approximate solution of the complex equation over a large domain. This procedure can be easily realized using commercial software COMSOL Multiphysics. Derivation of the proper equations for COMSOL Multiphysics solver which will be used to solve LL equation in linear approach complemented with Maxwell equations is presented below.

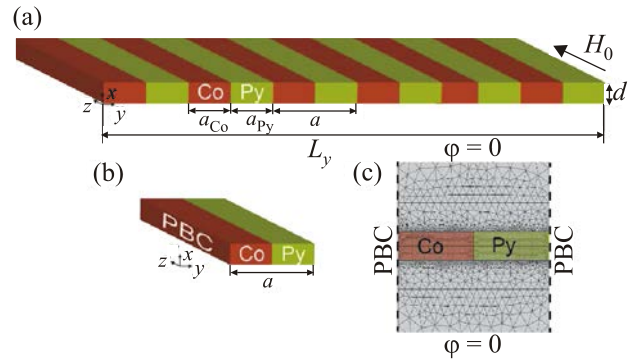

*Fig. 5.* (Color online) (a) 1D MC composed of parallel alternately ordered, connected witch each other Co and Py stripes (the stripes have thickness $d$ and width $a_{\text{Co}}$ and $a_{\text{Py}}$; lattice constant is $a = a_{\text{Py}} + a_{\text{Co}}$). The stripes are magnetically saturated along the z-axis by the external magnetic field $H_0$. (b) The unit cell of the 1D MC used in FEM computations. Periodic Bloch boundary conditions (PBC) are assumed along the y axis. (c) The computational area used in FEM. This area extends to large distance along the x axis above and below of the MC. At the borders of this area the magnetostatic potential takes the value 0.

To take into account appropriate electromagnetic boundary conditions for magnetostatic potential we considered MC which is surrounded by the nonmagnetic dielectric above and underneath the structure (Fig. 5 (c)), and assumed $\varphi = 0$ at the borders of the computational area, which are far from the MC. The magnetostatic potential shall fulfill Gauss Eq. (2) and because $\partial M_S / \partial z = 0$ we get

$$\nabla^2 \varphi = \frac{\partial m_x}{\partial x} + \frac{\partial m_y}{\partial y}, \qquad (11)$$

and from Eq. (3) we get

$$i\Omega \begin{bmatrix} m_x \\ m_y \end{bmatrix} = \begin{bmatrix} H_0 m_y - M_S \left( h_y + H_{ex,y} \right) \\ M_S \left( h_x + H_{ex,x} \right) - H_0 m_x \end{bmatrix}. \qquad (12)$$

We take here advantage of the Bloch theorem and assumed periodic boundary conditions (PBC) along the *y* direction. PBC are applied on the boundaries of the unit cell composed of Co and Py stripes, which is shown in Fig. 5(b). Solutions of Eqs. (11) and (12), according to Bloch's theorem for dynamic components of the magnetization, are in the form of Eq. (4) with $\mathbf{m_k}$ being periodic functions of *y* and dependent on *x* (i.e. across the thickness of the structure). Similar form for magnetostatic potential is taken:

$$\varphi(x, y) = \tilde{\varphi}(x, y) e^{i k_y y}, \qquad (13)$$

where $\tilde{\varphi}$ is a periodic function of *y* and depends also on *x*. $k_y$ is a Bloch wavenumber, which can be limited to the first Brillouin zone, i.e., to the range from $-\pi/a$ to $\pi/a$. Due to symmetry of the structure, this range can be further reduced to $k \in (0, \pi/a)$. Substituting Bloch functions of $\mathbf{m}$ and $\varphi$ (Eqs. (4) and (13)) to the system of Eqs. (11)–(12) we obtain the eigenvalue problem which is solved with the use of COMSOL 4.3a.

According to literature, we used the exchange field in the form appropriate for FEM calculations [46,49]:

$$\mathbf{H}_{ex} = \left[ \frac{1}{\mu_0 M_S} \nabla \cdot \left( \frac{2A}{M_S} \right) \nabla \right] \mathbf{m}(\mathbf{r}).$$

This formulation of the exchange field superimposed on the interfaces between magnetic materials introduces continuity of the dynamic magnetizations $m_i$ and $\frac{A}{M_s} \frac{\partial m_i}{\partial x}$.

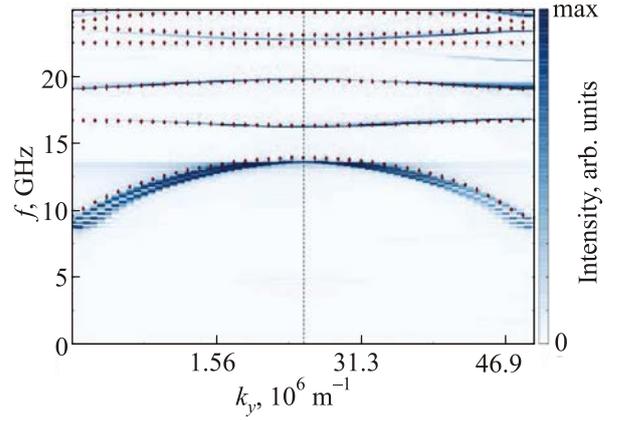

*Fig. 6.* (Color online) Dispersion relation of SWs in 1D MC. Blue lines correspond to the results obtained in micromagnetic simulations. Red diamonds mark the results obtained using FEM calculations in frequency domain. The vertical dashed line points the BZ boundary.

The solutions of Eqs. (11) and (12) satisfy also electromagnetic boundary conditions on the interfaces between magnetic materials and dielectric imposed by the Maxwell equations (i.e., tangential **H** component and normal **B** components are continuous).

Magnonic band structure calculated with FEM is shown in Fig. 6 with solid diamonds. The significant dispersion is visible for the first band, the second band is folded-back from the second BZ and has opposite slope. The third band has again positive slope. Between bands are magnonic band gaps. Width of the gap between 1st and 2nd band amounts to 3 GHz. This magnonic band structure is similar to the one measured by Brillouin light scattering experiment for 1D MC composed of Co and Py stripes but with lattice constant equal to 500 nm [22]. The identification of the SW excitations related to the bands can be made by the analysis of the SWs' amplitudes. These are obtained directly from FEM calculations as eigenvectors.

The spatial distribution of the selected modes on the $(x, y)$ plane is shown in Fig. 7. The 1st mode is a fundamental mode without any nodal points in the BZ center. For this mode, the SW amplitude is higher in Py than in Co. It results from the fact that the frequency range (around 9 GHz) of 1st band is well below FMR frequency of Co. It means that oscillations in Co stripes can be regarded as forced by the magnetization oscillations in Py. The modes with increased frequency have increased number of nodal points. For instance the 3rd mode which has two nodal points in the Py stripes, still don't have oscillations of the amplitude in Co. Interesting is also the 4th band. This mode has the nodal point across the thickness of the Py stripe (see, Fig. 7). This type of the excitations are named perpendicular standing spin waves (PSSW). Due to oscillations of the magnetization on short distance, exchange interactions determine their properties. The oscillations in neighbor Py stripes are not coupled via Co and the band does not change frequency with increasing $k_y$. However, for modes 1, 2 and 3 the change of the amplitude with increasing wavenumber is found. We can distinguish two kinds of changes with increasing Bloch wave vector $k_y$. First is associated with the change of the phase in the plane wave envelope of the Bloch function Eq. (4). At the BZ boundary (Fig. 7(c)) $k_y = \pi/a$ and the phase of the oscillations changes sign in the neighbor unit cells. Because harmonic oscillations appear only in Py, the zero of the SW amplitude in Co stripes for 1st and 3rd modes, and nonzero for the 2nd mode is observed at the BZ boundary.

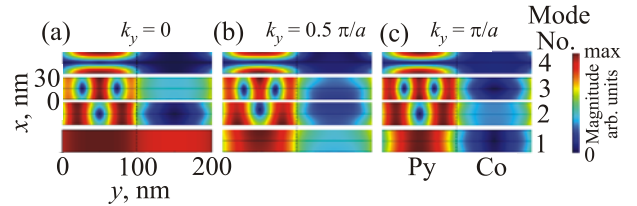

*Fig. 7.* (Color online) Distribution of the SW amplitude's modulus in the unit cell of the 1D MC composed of Co and Py stripes calculated with FEM. The amplitude for the 1st, 2nd, 3rd and 4th SW mode in the BZ center (a), for $k_y = 0.5\pi/a$ (b) and in the BZ border (c) of the dispersion relation from Fig. 6 is shown.

The second kind of change is caused by surface character of the DE wave [32]. In the homogeneous film the strength of the DE wave localization at the surface is proportional to the wavenumber. Thus, increase of the $k_y$ shall result in the surface localization of the SW also in MCs. Indeed, we observe asymmetric distribution of the SW amplitude across the thickness of the MC in the middle of the first BZ (Fig. 7(b)). The amplitude is larger at the top surface for the 1st and 3rd mode, and at the bottom surface for the 2nd mode. This distribution of the amplitude linked results of the periodicity and non-reciprocity of the DE wave. The DE wave propagating in the positive $y$ direction (for $H_0$ field pointed at positive $z$ axis) is localized at the top surface of the film, whereas the wave propagating in opposite direction is localized at the bottom surface. 2nd mode in the first BZ is obtained from the mode with $k_y = -0.75\pi/a$ (in the second BZ along negative direction of the wave vector) translated with reciprocal lattice vector $G = 2\pi/a$ to the first BZ, thus this mode it is localized at the bottom surface of the MC. 3rd mode is obtained from wavenumber $k_y = 1.25\pi/a$ (3rd BZ) by translation with $G = -2\pi/a$ to the first BZ. Thus, it is localized at the top surface and has larger localization than the 1st mode. The asymmetric distribution of the SW amplitude is lost at the BZ border, because at this wavenumber the superposition of the two counter-propagating waves is formed.

The results of the frequency domain FEM calculations will be verified by micromagnetic simulations in the following section.

**Micromagnetic simulations in time and real space domain**

Micromagnetic simulations (MMS) are very effective computational techniques extensively used for calculations of the magnetization dynamics in nano- and micro-sized ferromagnetic structures [33–39]. MMS are designed for solving numerically full LL Eq. (1) in the time domain and real space. There are many different MMS computational environments. They usually use one from the following numerical methods: Finite Difference Method (FDM) or FEM. Here, we use GPU accelerated MuMax3 software which is based on FDM with cuboidal space discretization [35].

Analysis of the SWs dynamics in time domain is very complex and time consuming task. The ideal situation is a single run of simulation which gives the full information about SWs dynamics and can be used to generate dispersion relation. However, in many cases, the results of several MMS are needed to obtain the dispersion relation of sufficiently fine resolution in frequency and wave vector domain, containing as many as possible branches of SW modes. In this section, the procedure for generation dispersion relation, based on MMS, will be described. That process will be exemplified by generation of the dispersion in 1D MC (Fig. 5(a)). The dispersion relation for this system, evaluated using FEM in frequency domain, was already discussed in the previous subsection.

MMS are performed in two stages. During the first stage, which can be called *static simulations*, equilibrium magnetic configuration is obtained. Then, this configuration is used as a starting point of the second stage, called *dynamic simulations*. During this step the static magnetic configuration is disturbed by adding relatively small (to stay in the linear regime and do not destroy equilibrium configuration) dynamic magnetic field. The dynamic field is usually directed orthogonally to the effective magnetic field. This perturbation of the magnetic field in turn induces coherent magnetization precession around the equilibrium direction. The form of dynamic magnetic field determines character of the excited SWs. It is important to use dynamic field which will excite as many SWs modes of different symmetries as possible, to collect information about full spectra of the SWs excitations. There are many possibilities of SWs excitations [40]. Very useful for that purpose are excitations in form of the sinc function, because sinc function is transformed in Fourier space to a window function. Here, we use dynamic magnetic field in form of product of two sinc functions (in space and time domain):

$$\mathbf{b}_{\mathrm{dyn}}(t, x, y, z) =$$
$$= b_0 \mathrm{sinc}\left[k_{\mathrm{cut}}(y - y_0)\right] \mathrm{sinc}\left[2\pi f_{\mathrm{cut}}(t - t_0)\right]\hat{n}, \quad (14)$$

where $b_0$ is the maximal value of magnitude appearing at time $t_0$ and at point $y_0$ along the periodicity direction, $\hat{n}$ is the unit vector perpendicular to the static magnetization. The $b_0$ should be small enough to stay in linear regime.

Space discretization of the MC and time steps for magnetization dynamics, as well as total size of the structure and total time of simulations, determine the correctness of MMS and the resolution of magnonic band structure. However, the requirement of the high resolution shall be compromised with the size of the computational problem to handle it with available computer resourcers. Maximal value of the frequency taken into account in MMS is $f_{\max} = 1/(2\Delta t_{\text{samp}})$ and maximal value of the wave vector is $k_{\max} = \pi/\Delta y_{\text{samp}}$, where $\Delta t_{\text{samp}}$ is an interval in time sampling and $\Delta y_{\text{samp}}$ is an interval in space sampling. Resolution of the dispersion relation (in the frequency and wave vector space) is determined by number of samplings taken into account: $\Delta f = 2 f_{\max} \Delta t_{\text{samp}} / t_{\text{sim}}$ and $\Delta k = 2 k_{\max} \Delta y_{\text{samp}} / L_y$, where $\Delta f$ and $\Delta k$ are frequency and wavenumber steps in dispersion relation, $t_{\text{sim}}$ is total time of the simulation and $L_y$ is total length of the sample. It shows that in many cases sampling intervals can be larger than space and time steps used by solver to space discretization and in time integration, respectively. Especially it is useful in the case of Runge–Kutta (RK) solvers [41]. For example, if we investigate SWs propagation up to 25 GHz, then optimal sampling frequency is $\Delta t_{\text{sampl}} = 2\cdot10^{-11}$ s. However, such time steps can be too long in many cases and it may be better to use shorter ones. It is convenient to use maximal step equal to $\Delta t_{\text{sampl}}$ and minimal step a few orders of magnitude shorter. Similarly is in the case of space discretization, but here, due to technical reasons, resampling is usually not convenient. Another important parameter in simulations is a number of periods in studied structure. This number of repetitions defines resolution of the dispersion relation in the wave vector domain (number of periods is the exact number of different **k** vectors inside 1st BZ). Summarizing, to obtain fine resolution of dispersion relation, it is necessary to simulate the system of many periods through long time.

The evolution of the magnetization in the whole system, obtained using MMS, can be written in the form of 4 dimensional matrix $\mathbf{M}(t,x,y,z)$. To reduce size of the problem, it is convenient to choose one component of the magnetization which is perpendicular to static magnetization — $M^c(t,x,y,z)$ where $c$ denotes chosen component of the magnetization, here we use $y$ component. To obtain dispersion relation for waves propagating along $y$ axis for certain values of $x_i$ and $z_i$ we calculate two-dimensional Fourier transform from $y$ and $t$ coordinates to $k_y$ and $f$. This is effectively realized with Fast Fourier Transform (FFT) algorithm $\mathcal{F}_{t,y}\{M^c(t,x,y,z)\} = \tilde{M}^c(f,x,k_y,z)$. Then, if we plot module of the obtained result $|\tilde{M}^c(f,x_i,k_y,z_i)|$ we will get the dispersion relation. If we aren't interested in modes visualization we can simply accelerate that process by choosing values $x_i$ and $z_i$ before calculating FFT: $\tilde{M}^c(f,x_i,k_y,z_i) = \mathcal{F}_{t,y}\{M^c(t,x_i,y,z_i)\}$. Results obtained in such way depend strongly on the particular form of the excitation. Note that, the different lines in dispersion relation have unlike intensities. Due to that, the dispersion relation could be unclear. Very helpful in extracting results from simulations are methods used for signal and image processing. More specific information about technical details can be found in Ref. 40.

Modes related with dispersion relation can be quite easily visualized. Firstly, the values of frequency ($f_0$) and wave number ($k_y$), corresponding to particular mode, should be selected. In next step we can reduce the size of obtained matrix $\tilde{M}^c(f,x,k_y,z)$ by leaving only elements corresponding to the selected frequency, $\tilde{M}^c(f_0,x,k_y,z) = \tilde{M}^c_{f_0}(x,k_y,z)$. Subsequently, the matrix is filtered, i.e. all values corresponding to different wave numbers than $k_0 + nG$, where $n \in \mathbb{Z}$ and $G$ is a lattice vector, are replaced by zeros: $\delta_{k_0+nG}\tilde{M}^c_{f_0}(x,k_y,z) = \tilde{M}'^c_{f_0}(x,k_y,z)$. In the further step one-dimensional inverse FFT is performed, separately for every value of $x$ and $z$: $M_{f_0,k_0}(x,y,z) = \mathcal{F}^{-1}_{k_y}\{\tilde{M}'^c_{f_0}(x,k_y,z)\}$. Then, the real (imaginary) part of that matrix, $\mathrm{Re}[M_{f_0,k_0}(x,y,z)]$ ($\mathrm{Im}[M_{f_0,k_0}(x,y,z)]$), can be plotted as profile of the particular mode corresponding to the points $(f_0,k_0)$ in $f(k)$ space belonging to the dispersion branch. Algorithm of calculating the dispersion relation and visualization of the particular modes is presented in Fig. 8.

To obtain dispersion relation already calculated with FEM (Fig. 6) we used 128 alternately ordered wires (Fig. 5(a)), which gives the total length $L_y = 12.8$ μm. We discretized this structure to 2600×1×4 cuboidal cells of sizes 5×20×7.5 nm. Due to symmetry along wires axis, it is possible to reduce number of cells along the $z$-axis during MMS by applying periodic boundary conditions.

We've intended to study SWs propagation for frequencies up to 25 GHz, therefore time of sampling intervals was set on $\Delta t_{\text{sampl}} = 2 \cdot 10^{-11}$ s. For SWs excitation we used small, orthogonal to static magnetic field, external dynamic magnetic field defined in Eq. (14), where cutoff frequency was $f_{\text{cut}} = 30$ GHz and cutoff wave vector $k_{\text{cut}} = 10^8$ m$^{-1}$. The parameters which shift value of the maximal excitation field in space and time was taken as $y_0 = L_y/2$ and $t_0 = 1 \cdot 10^{-9}$ s. To avoid nonlinear effects, the maximal amplitude was set as $b_0 = 0.05$ T. During dynamic simulations LL equation was solved using RK45 (Dormand–Prince) method [35] with maximal time step $\Delta t_{\text{sampl}}$ and minimal time step equal to $10^{-18}$ s. Total simulation time was set as 100 ns. The dispersion relation obtained in MMS is shown in Fig. 6 with the blue color map. Good agreement with results of FEM in frequency domain is obtained. However, the intensity of the different bands is different. It is related to the shape of the excitation field and its relation to the SWs amplitude [42,43]. The intensity of the 1st band is high, as this band is connected to the fundamental oscillations (Fig. 9) according also with FEM results (Fig. 7(a)).

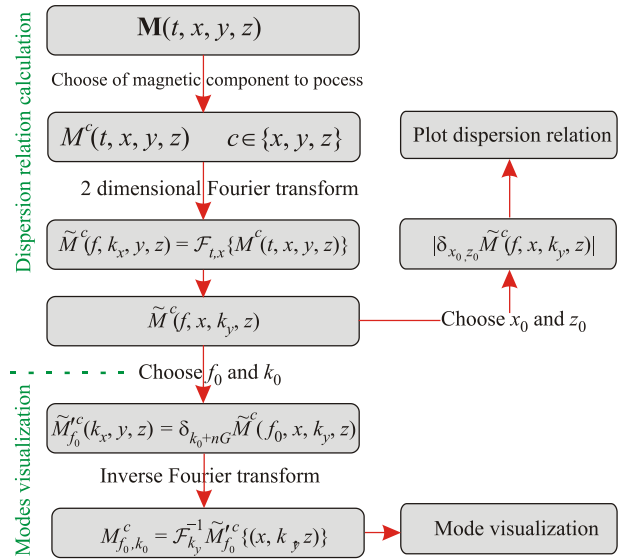

*Fig. 8.* (Color online) The post-processing algorithm of the MMS results in order to obtain dispersion relation of SWs in MC and to visualize amplitude of the SWs excitations.

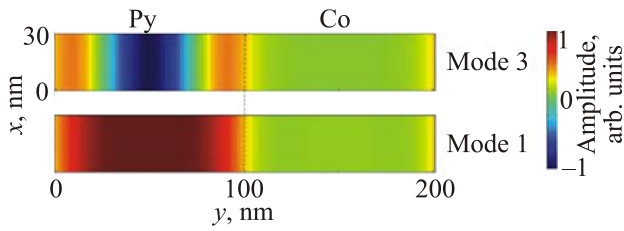

*Fig. 9.* (Color online) Distribution of the SW amplitude in the unit cell of the 1D MC calculated with MMS. The amplitude for the 1st, and 3rd mode in the BZ center is shown. Good match with amplitudes obtained with FEM is shown [Fig. 7(a)].

The 2nd mode is invisible in the dispersion relation at $k_y = 0$ obtained with MMS. It is because, for the symmetric excitation field, used in simulations (Eq. (14)), the asymmetric SWs in Py (see Fig. 7(a)) are not excited. However, with increasing wave number, the intensity of the second band in the dispersion relation increases. This is, because at the BZ boundary the oscillations are in opposite phase in neighboring unit cells.

### Nonreciprocity in magnonic crystals

In the previous two sections dispersion relation and amplitude of SW in 1D MC in the DE geometry were investigated. SWs propagating in thin homogeneous films, perpendicular to the external in-plane magnetic field (DE geometry), possess nonreciprocal properties discussed already in the previous section. This means that the SWs propagating in opposite directions (in $k_y$ or $-k_y$ direction) have amplitude of the dynamic magnetic fields distributed non-symmetrically across the film thickness and they have maximum near one of the surfaces of the film (maxima appear at opposite surfaces for $\pm k_y$) [18]. The localization of the amplitude has already been observed in the results of FEM calculations for 1D MCs (Fig. 7(b)). However, the frequencies of the oppositely propagating waves are the same, $f(k_y) = f(-k_y)$ (see Fig. 9).

Placing a metal overlayer atop of the film breaks spatial symmetry and also results in breaking symmetry of the dispersion relation with the respect to the change of the wavenumber sign [44]. The dispersion relation reveals significant change of frequency only for the spin waves propagating in one direction, resulting in nonreciprocal dispersion in the film $f(k_y) \neq f(-k_y)$. In other words, placing the metal overlayer causes spin waves with equal frequency propagating in converse direction to possess different wave vector magnitudes.

The effect of non-reciprocity has impact on the dispersion of MCs and the magnonic band gaps. Placing a metal plate or perfect electric conductor (PEC) on the top of magnonic crystal might leads to destruction of the Bragg condition, since the wavelengths of incident and reflected spin waves at fixed frequency are different [45]. However, the magnonic band gap can still form in this structure, as has been recently demonstrated theoretically and experimentally [46,47]. The spin waves propagating in opposite directions and possessing different values of wave vector magnitudes can interact and fulfill the general Bragg condition required for band gap opening.

FEM in the frequency domain has been implemented here to solve the LL and Maxwell equations in the magnetostatic approximation (i.e., neglecting dynamical coupling of the magnetostatic field with the electric field), to find the dynamical components of the magnetization vector **m**(**r**) and to obtain the dispersion relation in 1D MC shown in Fig. 10(a). The unit cell used in the calculation is shown in Fig. 10(b). The equations are defined by Eqs. (11) and (12). Because of PEC on the top surface we need to modify electromagnetic boundary condition at this surface. Here, the continuity of the normal component of the magnetic induction $B_x$ is required. We set the $B_x$ to zero on the top boundary of MC with PEC:

$$H_x = -\frac{\partial \varphi}{\partial x} + m_x = 0.$$

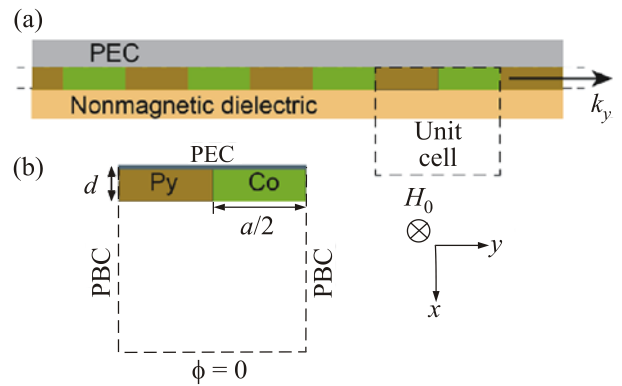

*Fig. 10.* (Color online) (a) A structure of 1D MC with a layer of a perfect electric conductor (PEC) on the top surface. The MC is composed of alternating, infinitely long stripes of Py and Co. The external static magnetic field $H_0$ is applied in the plane of the film, parallel to stripes. The SW propagate along $y$-axis. (b) The rectangular unit cell used in numerical calculations with periodic boundary conditions (PBC) assumed along the $y$ axis. The effect of a PEC being in the direct contact with the MC is implemented via boundary condition at the top surface. The bottom border of the unit cell is far from the MC.

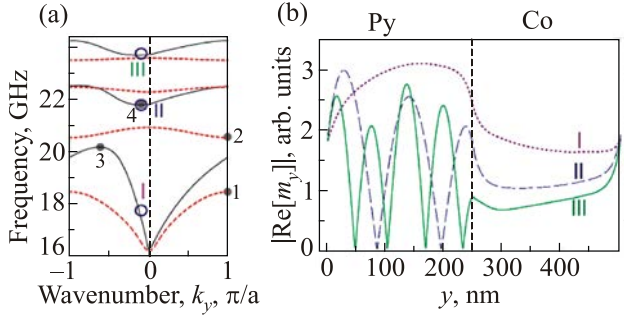

*Fig. 11.* (Color online) (a) The dispersion relation of the 1D MC composed of alternating Co and Py stripes with one side metalized (solid lines) and with both dielectric surroundings (dashed lines) for the bias field of $\mu_0 H_0 = 0.2$ T. (b) The absolute value of the dynamic magnetization |Re [$m_y(y)$]| of the first three modes: I, II and III of the Co/Py metalized structure for the wavenumbers marked in (a) with circles.

The dispersion calculated for MC with Co and Py stripes of width 250 nm, with dielectric surroundings is shown in Fig. 11(a) with dashed lines. Material parameters were taken from Refs. 48, 49 and the external magnetic field of 0.2 T was assumed. The dispersion is in good agreement with the experimental data from Ref. 23. The edges of the magnonic band gaps are indicated by points 1 and 2 at the BZ border in Fig. 11(a). This SW dispersion is reciprocal.

The nonreciprocity is introduced with the asymmetric boundary conditions between the top and the bottom surface of the MC. The magnonic band structure calculated for the same MC but with PEC at the top surface is shown by solid lines in Fig. 11(a). The dispersion has now nonreciprocal character and the edges of the band gap are shifted towards the center of BZ (points 3 and 4 in Fig. 11(a)). The band structure possess an indirect band gap. The Fig. 11(b) shows the distribution of $|\text{Re}[m_y(\mathbf{r})]|$ along the y-axis. The profiles of the dynamic magnetization components are laterally quantized for higher bands, showing that effective wave vector parameter increases with the number of mode.

### Magnonic band gap in three-dimensional magnonic crystals

In the last part we present the theory of 3D MCs. Concerning bi-component 3D MCs the big challenge is their fabrication, especially if the lattice constant is in the nanometer range. Generally, there are two approaches to the fabrication of such structures: top-down and bottom-up [50]. In top-down methods holes are drilled in the bulk material (matrix) and filled with another magnetic material (inclusions). Such methods are common rather in 2D case. The idea of bottom-up techniques is to prepare the lattice of inclusions and then to fill empty spaces with the matrix. This idea gives possibility to make 3D MCs using self-assembling magnetic nanoparticles (NPs) as a template. One of the most extensively studied example of magnetic NPs is magnetoferritin (mFT), a biomimetic NP based on a ferritin, a protein used in living organisms to store an iron in a nontoxic form [51,52].

The usage of cage-like proteins to grow magnetic NPs has a number of advantages [53] (extremely high level of homogeneity, variety of the sizes and properties of protein cages, diversity in physical or chemical functionality of protein shells) which allows to control the self-assembly process without modifying the NPs obtained inside the protein cages. Especially, mFT NPs can be filled with numerous magnetic materials resulting in different magnetic properties of the NPs [54,55]. The protein crystallization technique used to crystallize mFT NPs, allows to produce highly ordered 3D structures up to about 0.4 mm in size [56]. Obtained mFT crystals have high quality fcc structure and the lattice constant about 18.5 nm. An interesting effect is a reduction of the lattice constant to ca. 14 nm as a result of dehydration [57]. Moreover, it was shown theoretically that dried mFT crystals have the crystallographic structure and the lattice constant almost optimized for the occurrence of a complete magnonic band gap [58].

The object of this part of our study is a bi-component MC based on mFT crystal, i.e., a crystal consisting of mFT NPs (inclusions) arranged periodically in a ferromagnetic host material (B). The geometry of such MC is limited to fcc structure, in which mFT NPs crystallize, and the diameter of the inclusions is fixed at 8 nm (the diameter of the magnetic core of fully loaded mFT). The minimal lattice constant of such MC is 11.314 nm (mFT cores are touching each other). Constituent materials are characterized by magnetic parameters: the saturation magnetization $M_S$ and the exchange length $l_{ex}$. The contrast for each parameter is defined as the ratio of its value in mFT to its value in the matrix. An external magnetic field strong enough ($\mu_0 H_0 = 0.1$ T) to saturate the system is applied along one of the main crystal axes. The studied system is shown schematically in Fig. 12(c).

To determine the spin-wave spectra of 3D MCs we use PWM already introduced in Sec. 3 for 2D MCs. The magnetostatic field can be expanded in Fourier series using Eq. (2), as it was done for 2D MCs:

$$h_x(\mathbf{r}) = -\sum_{\mathbf{G}} \frac{(k_x+G_x)^2 m_{x,\mathbf{k}}(\mathbf{G}) + (k_x+G_x)(k_y+G_y) m_{y,\mathbf{k}}(\mathbf{G})}{|\mathbf{k}+\mathbf{G}|^2} e^{i(\mathbf{k}+\mathbf{G})\cdot \mathbf{r}},$$

$$h_y(\mathbf{r}) = -\sum_{\mathbf{G}} \frac{(k_x+G_x)^2 m_{y,\mathbf{k}}(\mathbf{G}) + (k_x+G_x)(k_y+G_y) m_{x,\mathbf{k}}(\mathbf{G})}{|\mathbf{k}+\mathbf{G}|^2} e^{i(\mathbf{k}+\mathbf{G})\cdot \mathbf{r}},$$

$$H_z(\mathbf{r}) \equiv H_{\text{dm}}(\mathbf{r}) = -\sum_{\mathbf{G}} \frac{M_S(\mathbf{G}) G_z^2}{\mathbf{G}^2} e^{i\mathbf{G}\cdot \mathbf{r}},$$

where position vector, the wave vector and reciprocal lattice vectors have now three components $\mathbf{r}=(x,y,z)$, $\mathbf{k}=(k_x,k_y,k_z)$ and $\mathbf{G}=(G_x,G_y,G_z)$ and the whole system is periodic in all three dimensions.

In a consequence also elements of the matrix (10) in the eigenvalue problem (9) are modified:

$$M_{ij}^{xx} = -M_{ij}^{yy}, \quad (15)$$

$$M_{ij}^{xx} = \frac{(k_x+G_{x,j})(k_y+G_{y,j})}{|\mathbf{k}+\mathbf{G}_j|^2} M_S(\mathbf{G}_i-\mathbf{G}_j), \quad (16)$$

$$M_{ij}^{xy} = H_0 \delta_{ij} + \sum_l (\mathbf{k}+\mathbf{G}_j)\cdot(\mathbf{k}+\mathbf{G}_l) l_{\text{ex}}^2 (\mathbf{G}_l-\mathbf{G}_j) M_S(\mathbf{G}_i-\mathbf{G}_j) +$$
$$+ \frac{(k_y+G_{y,j})^2}{|\mathbf{k}+\mathbf{G}_j|^2} M_S(\mathbf{G}_i-\mathbf{G}_j) - \frac{(G_{z,i}+G_{z,j})^2}{|\mathbf{G}_i+\mathbf{G}_j|^2} M_S(\mathbf{G}_i-\mathbf{G}_j),$$

(17)

$$M_{ij}^{yx} = -H_0 \delta_{ij} - \sum_l (\mathbf{k}+\mathbf{G}_j)\cdot(\mathbf{k}+\mathbf{G}_l) l_{\text{ex}}^2 (\mathbf{G}_l-\mathbf{G}_j) M_S(\mathbf{G}_i-\mathbf{G}_j) -$$
$$- \frac{(k_x+G_{x,j})^2}{|\mathbf{k}+\mathbf{G}_j|^2} M_S(\mathbf{G}_i-\mathbf{G}_j) + \frac{(G_{z,i}+G_{z,j})^2}{|\mathbf{G}_i+\mathbf{G}_j|^2} M_S(\mathbf{G}_i-\mathbf{G}_j).$$

(18)

By diagonalization of the matrix (10) with sub-matrices defined in Eqs. (15)–(18), we obtained the reduced frequencies $\Omega$ (eigenvalues) and the eigenvectors — coefficients of the Bloch expansion of the dynamic part of the magnetization (5), as in 2D MC.

In Fig. 12 we present two spin-wave spectra plotted along the line connecting high-symmetry points in the first BZ. The lattice constants is assumed to be 18.5 nm (this value correspond to mFT crystal as prepared) and the magnetic parameters of the mFT NPs are: $M_S = 0.346\cdot 10^6$ A/m and $A = 10^{-11}$ J/m [59]. Two matrix materials are used: Co in (a) or Ni in (b). For Co matrix a wide (ca. 100 GHz), complete magnonic band gap appears. This gap is located between the first and the second band. In the case of Ni matrix the band gap failed to open which is the consequence of too weak contrast of saturation magnetization $M_S$ [60]. The existence of a critical $M_S$ contrast below which the complete gap does not open stems from the fact that in a spectrum for 3D structure a complete bandgap is *indirect* gap. The bottom and the top of this gap is related to different propagation directions (in contrast to the *directional* gap, i.e., the gap for particular direction of the propagation). Roughly speaking, the width of the directional gap depends on the contrast of magnetic parameters, and its central frequency is determined by the value of the wave vector at the boundary of the first BZ. If directional gaps for different directions of propagation are too narrow, then the overlapping of neighboring bands appears. This very effect underlies the lack of complete magnonic gap for the Ni matrix (Fig. 12(b)).

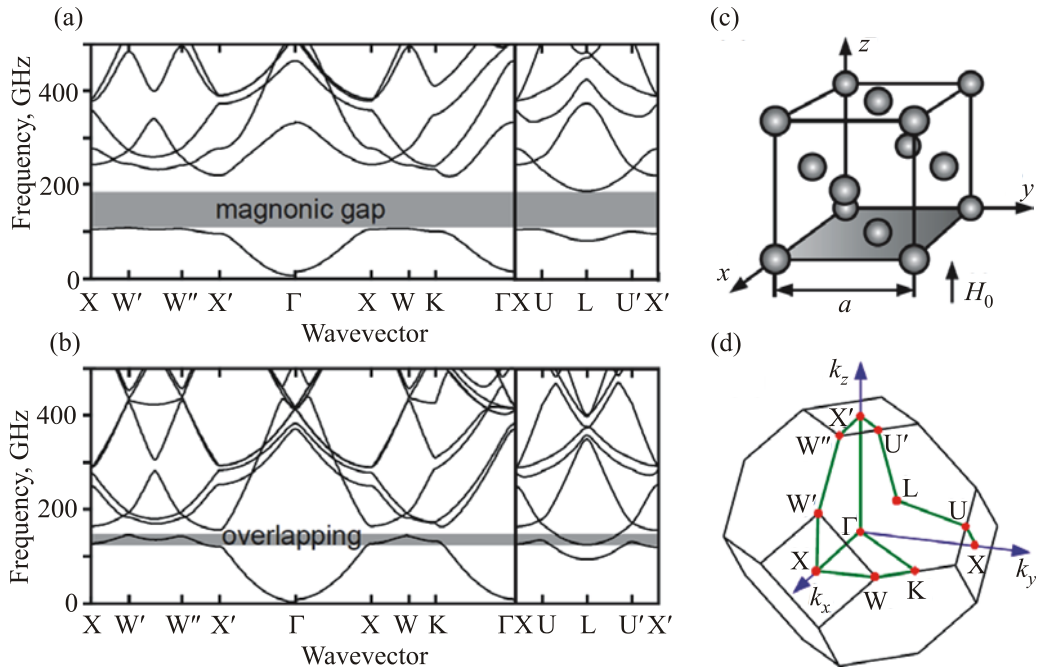

*Fig. 12.* Spin-wave spectra (up to 500 GHz) of mFT MC with (a) cobalt and (b) nickel matrix for lattice constant 18.5 nm (mFT crystal as prepared). The spectra are plotted along the high-symmetry path in the 1st Brillouin zone shown in (d). Shaded area represents: in (a) the complete magnonic band gap and in (b) the overlapping of the first and second band. (c) Schematic depiction of the mFT MC structure with coordinating system used.

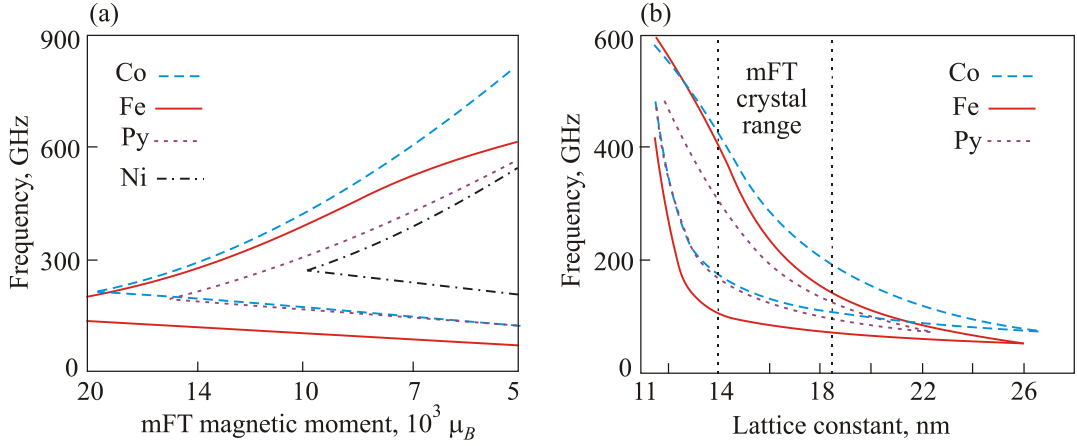

*Fig. 13.* (Color online) The edges of the complete magnonic gap vs. (a) magnetic moment of inclusions and (b) lattice constant of MC for a Fe, Co, Py and Ni matrix. In (a) the lattice constant of MC is fixed at 14 nm (dried mFT crystal). In (b) magnetic moment of inclusions is fixed at the typical value for mFT NPs ($10^4$ $\mu_B$) and there is no gap for Ni matrix in this case.

As we have already mentioned, the mFT NPs can be loaded with different magnetic materials which means that the contrast of $M_S$ can be modified in quite broad range. In Fig. 13(a) we show the evolution of the complete magnonic gap vs. the total magnetic moment of mFT NPs in MCs, which are based on four matrix materials: Fe, Co, Py and Ni. The lattice constant is fixed to 14 nm (dried mFT crystal). Presented dependences range from the magnetic moment twice the typical value of mFT ($10^4$ $\mu_B$) [57] which is gradually reduced to half of this value. The gap widens quickly with increasing $M_S$ contrast (decreasing magnetic moment of the mFT NPs), consequently the complete magnonic gap opens even for Ni matrix at $9.9 \cdot 10^3$ (a value only 1% less than in typical mFT NPs).

Another way to tailor the magnonic gap is adjusting to the MC lattice constant. As we already mentioned, the mFT crystal reduces its lattice constant from 18.5 to 14 nm while drying. Additionally, functionalization of the external surface of protein cage is also promising feature in terms of controlling the lattice constant of the MC. We plotted the evolution of edges of the complete magnonic gap vs. lattice constant for different matrix materials (Co, Fe, and Py) to explore the ranges in which the complete magnonic gap exists (see Fig. 13(b)). The magnetic moment of inclusions is fixed to $10^4$ $\mu_B$. In this case there is no gap for Ni matrix for any value of the lattice constant. For all cases we studied, the complete magnonic gap changes a lot with the lattice constant and its maximal width is observed at approximately 13 nm (the precise value depends on the matrix material).

The existence of this maximum is related to the concurrence of the exchange and dipolar interactions. Coexistence of long- and short-range interactions leads to very interesting phenomena in variety of systems [61–64]. In the MCs concerned in this study it strongly influences the spin wave profiles of two lowest modes (Fig. 14). For small lattice constant the modes of lowest frequency have a bulk character due to the strong exchange interactions between excitations in neighboring areas of the MC. As the lattice constant grows, the significance of exchange interactions fades while dipolar interactions gain in importance. As a result for the larger lattice constant, stronger spin wave profiles are concentrated in mFT NPs (first mode) or in the matrix (second mode). For in-between region, where both types of interactions have comparable weightness, maximum of the gap width appears.

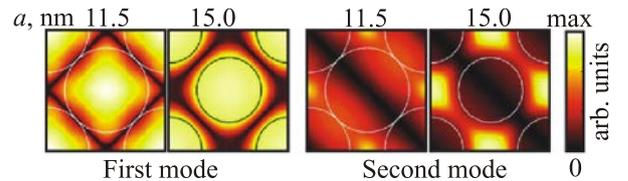

*Fig. 14.* (Color online) Profiles of dynamic magnetization for the two lowest spin-wave modes in an mFT/Co MC in a plane perpendicular to the external field and passing through the centers of mFT NPs (the contours of which are represented by circles).

The profiles of the spin waves have great influence on the effective damping as well. Using PWM based on LL equation with damping included [65], we found the concentration of the mode in one of the constituent material changes effective damping towards its value in this material [65]. This issue can lead to strong anisotropy of the effective damping. As an example in Fig. 15 we show results for the lowest modes obtained in MC containing spherical scattering centers (material B, $M_{S,B} = 0.194 \cdot 10^6$ A/m, $A_B = 3.996 \cdot 10^{-12}$ J/m — these values are close to yttrium iron garnet) disposed in the matrix (material A, $M_{S,A} = 1.752 \cdot 10^6$ A/m, $A_A = 2.1 \cdot 10^{-11}$ J/m — the values close to Fe). The lattice is assumed to be simple cubic with the lattice constant $a = 10$ nm and the sphere radius equal to 3.628 nm. The Gilbert damping parameter was chosen as $\alpha_A = 0.0019$ and $\alpha_B = 0.064$. The eigenmodes being solutions of LL are characterized by complex eigenvalues $\Omega' = \Omega + i\Omega''$. In Fig. 15(a) and (b) we plotted the real and imaginary part of $\Omega$, respectively. The imaginary part $\Omega'$ is a measure of the life time of the particular SW excitation. However, the value $\Omega''$ has to be referred to $\Omega'$ for direct comparison of the damping of different modes. In Fig. 15(c) we show so called "figure of merit" (FOM) which is the real part of the frequency divided by its imaginary part. As we can see the damping of the lowest mode is much higher (lower FOM) than for the second mode. To explain this feature we plotted spin-wave profiles (the distribution of the dynamical part of magnetization) for two lowest bands (see Fig. 15(d)). It is clearly seen that the lowest mode is strongly concentrated in scattering centers (with higher damping) while the second one — in the matrix (with lower damping). Therefore, the effective damping for these two modes is much different. For the same reason the effective damping depends on the propagation direction as well. For example: comparing the profiles of the lowest mode for two different wave vectors, the concentration of the dynamical magnetization in the spheres is a bit stronger for direction R than at point Γ. This results in the difference of effective damping: 0.059 for R and 0.045 for Γ.

### Conclusions

We have demonstrated the application of different numerical methods: PWM, FEM and MMS for calculations of the magnonic band structure in 1D, 2D and 3D MCs. The PWM is an useful tool for calculation of SW properties in periodic structures, independent on its dimensionality, shape of the elements and crystallographic lattice. However, PWM is limited to fully saturated materials and it works only in linear approximation. In the case of PWM applied to planar MCs, the considered MCs have to be homogeneous across the thickness and relatively thin. In the proposed FEM the last assumption is avoided. We showed usefulness of this method in the study of inhomogeneous excitations across the thickness or inhomogeneous film of MC. We showed that FEM can be easily extended to include the boundary effect induced by conductive materials surrounding MC film. This method is suitable for investigation of the nonreciprocal properties in SW dynamics in 1D and 2D MCs. MMS are the most general tool, which avoid assumptions used in PWM and FEM. However, this method requires the complex post-processing to extract information obtained directly in PWM or FEM. The other drawback of MMS is a long time of the calculations, however, this is cut down by implantation of the GPU units for calculations. We have demonstrated usefulness of MMS to study magnonic band structure in planar 1D MC. Its extension to the planar 2D MC can be easily done.

We have demonstrated, that the surface character of the Damon–Eshbach wave is also preserved in magnonic crystals. Surface localization increases with the band number, however the localization exists only inside the Brillouin zone, this is excluding Brillouin zone center and border. The surface property of the spin wave excitation is further exploited by covering plate of the magnonic crystal with a conductor. Due to that the nonreciprocal dispersion relation is introduced.

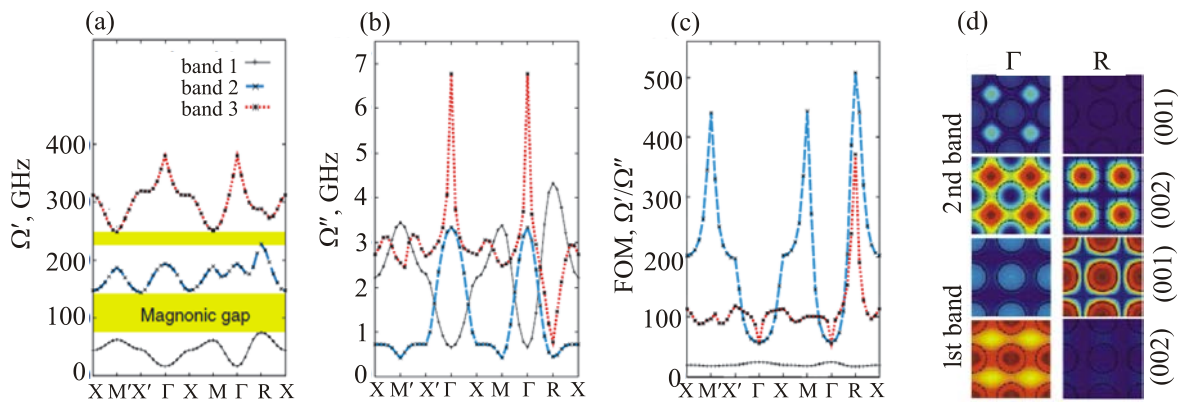

*Fig. 15.* (Color online) The real and imaginary parts of the frequency in the first BZ for sc MC in (a) and (b), respectively. (c) The figure of merit (FOM) for the same structure. (d) The distribution of the amplitude of the dynamical components of the magnetization across the planes perpendicular to the external field. Planes (001) and (002) are between and across the spheres, respectively. The profiles from the first and the second band in Γ and R point of the first BZ are shown. (Figures taken from Ref. 65).

The band structure in 2D magnonic crystals is very complex. It is caused not only by structurization, but also by the demagnetizing field. This field introduces additional spatial inhomogeneity for spin waves and modifies spin wave dispersion. It also makes the band structure strongly dependent on shape of inclusion and type of lattice for magnonic crystal. Pronounced effect is the localization of low frequency spin waves in the areas of the lowered internal magnetic field. However, the inhomogeneity of the internal magnetic field becomes unimportant for magnonic crystals with small lattice constant where demagnetizing effect can be neglected. In this limit, the band structure is only slightly dependent on the shape of inclusions and type of the lattice. For 3D magnonic crystals, characterized by small lattice constant, for the structures with characteristic sizes comparable to diameter of magnetoferritin crystals, the wide band gap was found. Finally, we have pointed out that the spatial distribution of different materials could be explored for tailored effective damping of spin waves.

### Acknowledgments

The research leading to these results has received funding from Polish National Science Centre project DEC-2-12/07/E/ST3/00538 and from the EUs Horizon2020 research and innovation programme under the Marie Sklodowska-Curie GA No644348. The numerical calculation were performed at Poznan Supercomputing and Networking Center (grant No. 209).